%% file: paper.tex
\documentclass[11pt]{article}

\usepackage{verbatim}
\usepackage{amsmath}
\usepackage{amsfonts}
\usepackage{latexsym}
\usepackage{epsfig}
\usepackage{color}

\input{texdefs}

\begin{document}

\input{title}

\setcounter{page}{1}

\input{intro}

\input{bosonic2}

\input{superfield}

\input{superstring}

\input{comparisons}


\input{concl}

\appendix 
\def\theequation{\thesection.\arabic{equation}}

\setcounter{equation}{0}
\input{appSums}

\bibliographystyle{unsrt}
{\small
\bibliography{paper}
}

\end{document}

%% file: texdefs.tex
\DeclareMathSymbol{\mg}{\mathrel}{symbols}{"1D}

\newcommand{\df}{\partial}
\newcommand{\ddf}{\bar \partial}
\newcommand{\beq}{\begin{equation}}
\newcommand{\eeq}{\end{equation}}
\newcommand{\barr}{\begin{array}}
\newcommand{\earr}{\end{array}}

\newcounter{oldcounter}

\newcommand{\ba}[2]{\[\begin{array}{#2}\label{#1}}
\newcommand{\ea}{\end{array}\]}
\newcommand{\be}{\begin{equation}}
\newcommand{\ee}{\end{equation}}
\newcommand{\bea}{\begin{eqnarray}}
\newcommand{\beqn}{\begin{eqnarray}}
\newcommand{\eea}{\end{eqnarray}}
\newcommand{\eeqn}{\end{eqnarray}}

%% file: title.tex
\thispagestyle{empty}

\begin{flushright}
\end{flushright}

\vskip 2 cm
\begin{center}

{\Large 
{\bf 
Comments on Higher Derivative Terms for the Tachyon Action
}
}

\vspace{1.23cm}
{
\large
{
\bf Mark Laidlaw$^{a,}$\footnote{ Email:  
\texttt{mlaidlaw@perimeterinstitute.ca}  }, 
{
\bf Madoka Nishimura$^{b,}$\footnote{
Email: \texttt{mnishimu@sciborg.uwaterloo.ca} }\footnote{Address after 
April 1:   Department of Theoretical Physics, Uppsala University, Box 803, 
SE-751 08 Uppsala, Sweden}, \bigskip 
}\\[0pt]
\vspace{0.23cm}
${}^a$ {\it  
Perimeter Institute for Theoretical Physics, \\
35 King Street North, Waterloo, Ontario N2J 2W9, Canada
}\\[0pt]
\vspace{0.23cm}
${}^b$ {\it Department of Physics, University of Waterloo, \\
Waterloo, 
Ontario N2L 3G1, Canada }
\bigskip
}}
\end{center}
\subsection*{\centering Abstract}

We consider the open string tachyon action in 
a world--sheet sigma model approach.
We present explicit calculations up 
to order $8$ in derivatives for the bosonic string, and mimic these to 
order $6$ for the superstring,  
 including terms with multiple 
derivatives acting on the tachyon field.  
We reproduce lower derivative terms 
obtained elsewhere, and speculate on the role of the world--sheet contact 
terms in regularizing the action for the superstring tachyon.

 \newpage

%% file: intro.tex
\section{Introduction}
\label{sec:intro}

The existence of a tachyon in a field theory is generally understood to be 
indicative of an instability in the theory; that the theory is being 
expanded around an incorrect vacuum.  There has therefore been a great 
deal of interest in investigating the dynamics associated with the 
tachyonic degrees of freedom that appear in the spectrum of both bosonic 
and superstring theories.  The condensation of the open string tachyon 
into a stable vacuum remains an interesting open problem addressed in
off--shell string theory.  

Some of the original work on this subject 
\cite{Witten:1992qy,Witten:1993cr,Shatashvili:1993kk,Shatashvili:1993ps}  
was concerned with obtaining an
action for the tachyon field by integrating out stringy degrees of freedom
to obtain a (string) partition function in which the couplings were 
related to the target space fields.   This partition function was to be 
interpreted as a space time action for the fields thus described, and the 
efforts were in parallel with the derivation of the Born--Infeld action 
from strings interacting with a background gauge field 
\cite{Fradkin:1985qd}.  The
tachyon profiles originally considered were manifestly off--shell.
Their particular form made the world--sheet theory  solvable, 
and exact expressions for the action were obtained in this regime 
\cite{Witten:1993cr,Kraus:2000nj,Craps:2001jp,Kutasov:2000qp}.  
This line of inquiry was generalized to the superstring 
\cite{Kutasov:2000aq} 
and it has been understood that even such simple  profiles were useful toy 
models in the study of D--brane decay 
\cite{Viswanathan:2001cs,Gerasimov:2000zp,Laidlaw:2001jt}.  A 
more recent approach 
to this has been to study the branes described by the `rolling tachyon' 
solution \cite{Sen:2002in,Sen:2002an}.  This solution is an exactly 
marginal perturbation on 
the string world--sheet boundary and describes the time--dependent decay 
of a brane.  It 
has been shown that, when coupled to a gauge field, the decay products of 
this process is a radiation of closed strings carrying the polarization of 
the coupled magnetic field \cite{Mukhopadhyay:2002en,Sen:2003bc}.

Invoking other considerations it has also been argued  that what is 
commonly 
called  the `tachyon DBI action',
\beqn
S = \int V(T) \sqrt{ \det \left( g_{\mu\nu} + \df_\mu T \df_\nu T + 
\df_\mu \Phi^i \df_\nu \Phi_i \right) }
\eeqn
reproduces the dynamics for the D--brane system with transverse scalars  
$\Phi_i$.  This action has been the subject of numerous studies, for 
example 
\cite{Kutasov:2003er,Sen:2002an,Garousi:2000tr,Quevedo:2002xw,Kluson:2004qy}, 
and describes an aspect of the tachyon 
interaction with other degrees of freedom, even in the off--shell case.  
The tachyon DBI action is also the Lorentz 
covariant generalization of an exact solution of the tachyon action from 
the rolling tachyon \cite{Larsen:2002wc}.  In an  attempt to 
better understand 
the off--shell dynamics of the string, it has been proposed to superimpose 
a 
spatial dependence on the rolling tachyon solution 
\cite{Fotopoulos:2003yt} and 
examine the corresponding action.  For certain (periodic) spatial 
dependences exact solutions have been found \cite{Rey:2003zj,Sen:2002qa}. 
The 
more frequent 
practice follows the calculation of derivative 
corrections to the Born--Infeld action \cite{Andreev:1988cb} using an 
expansion around 
a marginal boundary interaction. The action is understood to be 
accurate to a certain order in spatial derivatives.  

A different approach which is much less reliant on world--sheet properties 
is to extract on--shell matrix elements for different scattering 
processes \cite{Garousi:2003ai}.  Once these are calculated the 
appropriate terms in 
the action 
can then be reconstructed.  One difficulty with this approach is that the 
starting point is  an on--shell calculation, and  so is subject to 
ambiguities when one tries to 
extend to off--shell 
processes  (for a recent discussion, see \cite{Niarchos:2004rw}).

An interesting  paper \cite{Fotopoulos:2003yt} raised a number of 
points with 
respect to these lines of investigation:  The argument 
which established the tachyon DBI action as an exact solution for the 
rolling tachyon \cite{Larsen:2002wc} does not trivially extend to spatial 
variations, and so motivates a search for the Lorentz invariant 
generalization.  The second was to ask the question about the meaning of 
the tachyon action,  obtained by integrating out the stringy 
modes:  In the case of a massless particle the higher stringy modes are 
separated by a mass gap, however for the tachyon the (negative) mass 
squared is of the same order as the infinite tower which introduces 
subtleties to the 
interpretation of the effective action.  The third point was that the 
tachyon DBI action  depends only on first derivatives of the 
tachyon field.  It was argued that a consistent expansion should also 
contain higher order derivatives on the tachyon field, as obtained through 
integration by parts.  

We are motivated by the argument that the tachyon action 
should be Lorentz invariant and contain all possible combinations of 
derivatives, and in this paper we attempt to address these points
following the 
world--sheet techniques 
of \cite{Takayanagi:2000rz}.
We find that the technique has an 
obvious generalization to problems like the expansion around the rolling 
tachyon solution.  It also offers some lessons about the derivative 
structure that is `natural' from the world--sheet point of view:  
that for off--shell processes there is a  non--local field 
redefinition which depends on the normal ordering 
prescription for the sigma model.  Related to this, 
the  action does not have 
any terms which depend on the 
derivatives  $\df_\mu \df^\mu$ acting on a single field.
We are also motivated to understand the contact term in the boundary 
theory on the superstring theory.  This term is required for 
unbroken world--sheet supersymmetry \cite{Kutasov:2000aq}, but 
this term is omitted from the analysis
in 
\cite{Larsen:2002wc}. 
The 
omission was  
explained in \cite{Fotopoulos:2003yt} as being justified in the momentum 
region under consideration.  We speculate that this contact term is 
necessary to appropriately regularize the tachyon action in particular 
momentum regions.

This paper is organized as follows, in section \ref{sec:bosonic}
we  review the derivation of the action for the boundary tachyon 
coupled to a bosonic string.  
This will have an obvious generalization to the case of superstring 
theory, and will illustrate the mechanics of our calculation.  
In section
\ref{sec:superfield} we illustrate the calculation for the superstring 
using a superfield formalism, and find an action for the tachyon up to $6$ 
derivatives.  In section \ref{sec:concl} we summarize and discuss our 
results.  A number of the 
technical details, including the regularization prescription, are 
relegated to 
appendix \ref{sec:sumsappendix}.

%% file: bosonic2.tex
\section{Effective action for bosonic string tachyon}
\label{sec:bosonic}

Prior to performing the calculation for the superstring, we present a 
derivation of the derivative expansion for the bosonic open string tachyon 
action.  We do this to illustrate the technique used 
without the additional complication of fermion fields on the string 
world--sheet, and also because the origin of  some features that are 
reproduced in the 
superstring are easier to trace in this simplified setting.    

The method used here is inspired by the worldsheet approach of 
\cite{Fradkin:1985qd}.  
Specifically we integrate out the world--sheet degrees of freedom 
and obtain a partition functional of the tachyon field.  We expand 
to 
a fixed order in derivatives of the tachyon field and interpret the 
resulting expression as the first terms of a covariant expression  
for the tachyon action.

We introduce the tachyon field as a boundary interaction of the unit disk 
with the action
\bea
S &=& \frac{1}{2\alpha'} \int \frac{d^2 z}{2\pi} \df X^\mu \ddf X_\mu
+ \oint \frac{d\phi}{2\pi} T(X(\phi))
\label{eq:bosonicaction}.
\eea
In (\ref{eq:bosonicaction}) $\phi$ is the coordinate on the boundary of 
the world--sheet.  We then divide the field $X$ into a 
classical and oscillatory part as $X \rightarrow x + \tilde X$, and 
write the mode expansion as 
\beqn
T(X) = \int dk T(k) e^{i k x } e^{i k \tilde X}.
\eeqn
We now follow {\cite{Witten:1993cr,Tseytlin:2000mt}} and calculate a 
partition functional that depends 
on the tachyon by integrating out the oscillatory modes $\tilde X$ using 
the action (\ref{eq:bosonicaction}), explicitly  
\beqn
Z\left( T (x) \right) &=& \int d\tilde X e^{-S\left( T(x+\tilde X) 
\right)}.
\eeqn
Once we have obtained the partition functional for the bosonic string, we 
may calculate the space--time action as
\beqn
S \left( T (x) \right) &=& \left( \beta_{T} \partial_{T} +1 \right) 
Z\left( T (x) \right).
\eeqn
We 
expand $Z$ as a power series in $T$ 
and find that at order $T^n$ we have \beqn
\frac{(-1)^n}{n!} \oint \prod_{i=1}^n \frac{d\phi_i}{2\pi}
\langle \prod_{i=1}^n T\left( x+ \tilde X(\phi_i) \right) \rangle.
\eeqn
This can be evaluated as
\beqn 
\frac{(-1)^n}{n!} \oint \prod_{i=1}^n \frac{d\phi_i}{2\pi}
\left[ \prod_{i=1}^n \int dk_i e^{i k \cdot x} T(k_i) \right]
\prod_{i=1}^n e^{- \frac{k_i^2}{2} B(0) } 
\prod_{i<j=1}^n e^{- k_i \cdot k_j B(\phi_i - \phi_j) },
\label{eq:nthbosonicterm}
\eeqn
where $B(\phi-\phi')$ is the propagator for the oscillatory modes, given by
\beqn
B(\phi - \phi') &\equiv& 
\langle \tilde X^\mu (\phi) \tilde 
X^\nu (\phi') \rangle  
\label{eq:bosonicprop}
\\ \nonumber &=& -2 \alpha' 
g^{\mu\nu}
\ln \left| 2 \sin \frac{ \phi - \phi'}{2} \right| \nonumber \\
&=& 2 \alpha' g^{\mu\nu} \sum_{m = 1}^\infty \frac{ \cos m (\phi - \phi')
}{m} . 
\label{eq:bosonicpropagator}
\eeqn
Our convention for the metric is that the signature of $g^{\mu\nu}$ is 
$\left( -, + \ldots + \right)$. 
The expression (\ref{eq:nthbosonicterm})
is nothing but an off--shell bosonic string tachyon 
scattering 
amplitude for general values of $k$ and $T(k)$. Due to 
the off--shell nature of this calculation we have retained a term 
associated 
with self contractions of the tachyon vertex operators.  For the same 
reason we have not used conformal invariance to fix the location of any 
of the tachyon insertions.
We absorb the self contractions by a non--local field 
redefinition 
\beqn
\tilde T(x) &=& e^{c \df_\mu \df^\mu} T(x),
\label{eq:nonlocalredef}
\eeqn
where $c$ is a constant associated with the regularized version of  $B(0)$ 
and defined by 
$c=\frac{B(0)_{\mathrm{reg}}}{2}$.  This term appears in some other 
analyses, such as \cite{Coletti:2004ri, Fujita:2004ha}, and on--shell it 
is simply a rescaling, but off--shell it amounts to redefining the 
tachyon field to include the anomalous dimension.

We can obtain an exact result for the integral (\ref{eq:nthbosonicterm}) 
when we consider a tachyon background with a single momentum mode and  
profile $T_{p}(X) = T_0 e^{i p 
\cdot X}$.
In this case the integral is trivially recast as Dyson's integral, a 
special case of Selberg's generalization of the $\beta$--function 
\cite{book:specialfunction}.  Then we obtain for $n$ insertions of $T$ 
\beqn
\frac{(-1)^n}{n!} 
\left[ \tilde T_{p}(x) \right]^n
\oint&&\prod_{i=1}^n \frac{d\phi_i}{2\pi}
\prod_{i<j=1}^n e^{-  p^2 B(\phi_i - \phi_j) } \nonumber \\
=&& \frac{(-1)^n}{n!}
\left[ \tilde T_{p}(x) \right]^n
\oint \prod_{i=1}^n \frac{d\phi_i}{2\pi}
\prod_{i<j=1}^n \left| e^{i \phi_i} - e^{i \phi_j} \right|^{2 \alpha' p^2}
\nonumber \\
=&&
\frac{(-1)^n}{n!}
\left[ \tilde T_{p}(x) \right]^n
\frac{\Gamma(1 + n \alpha' p^2) }{ \Gamma(1+\alpha' p^2)^n }
\label{eq:bosonplanewave}.
\eeqn
This expression interpolates between two  known cases, the expansion 
around the constant tachyon which gives rise to a potential $e^{-T_{p}}$ 
\cite{Witten:1993cr}, 
and the expansion around the rolling tachyon which gives a potential 
$\frac{1}{1+ T_{p}}$ \cite{Larsen:2002wc}.  The correlation function 
(\ref{eq:bosonplanewave}) can also be expanded in perturbation around this 
plane wave solution using $p^2 \rightarrow p^2 + \delta p^2$.   The 
variation about the constant tachyon clearly vanishes at first order in 
$\delta p^2$.  About the $p^2=1$ rolling tachyon case it produces 
$\left[ \tilde T_{p}(x) \right]^n n \left( \psi(1+n)-\psi(2) \right) 
\delta p^2$, and re-casting the momentum as derivatives, and summing over 
$n$ reproduces the spatial derivative term of \cite{Fotopoulos:2003yt}.
The case $n=2$ of (\ref{eq:bosonplanewave}) can be immediately extended 
to the well--known 
expression for the two--point function for the bosonic tachyon 
\cite{Coletti:2004ri,Kutasov:2000qp}, 
\beqn
\frac{\tilde T^2}{2!}
 \frac{ \Gamma[1 + 2 \alpha' k_1 \cdot k_2] }{ \Gamma^2[1+ \alpha' k_1 
\cdot k_2]} 
\label{eq:grigt2}
\eeqn
and in \cite{Coletti:2004ri} the three--point function for the bosonic 
tachyon is derived to be 
\beqn \frac{\tilde T^3}{3!}
\Gamma( 1+k_{12}+k_{13}+k_{23} ) \prod_{i<j=1}^3 \frac{\Gamma( 1 + 
2k_{ij})}{ \Gamma( 1 + k_{ij} ) \Gamma( 1+k_{12}+k_{13}+k_{23} - k_{ij} )}
\label{eq:grigt3}
\eeqn
where $k_{ij} = \alpha' k_i \cdot k_j$.  (\ref{eq:grigt3}) is consistent 
with 
(\ref{eq:bosonplanewave}) in the case of a single momentum mode.

We wish to understand the application of these exact expressions  we 
consider the region around the constant tachyon field and our 
strategy is  to expand 
the expression (\ref{eq:nthbosonicterm}) in powers of $k$, do the 
integration over the insertion points for the field $T$ on the 
world--sheet boundary, and then to 
Fourier transform back to  position space.  This  
identifies the expansion in powers of $k$ with an expansion in powers of 
derivatives, and is 
similar to the strategy used in \cite{Andreev:1988cb} to obtain the 
Born--Infeld
action and derivative corrections from a world--sheet approach.  We note 
that do not expand around a conformal solution, because while $F_{\mu\nu} = 
\mathrm{constant}$ is on--shell, $T=\mathrm{constant}$ is not.  
Using standard field theory techniques and the propagator 
(\ref{eq:bosonicpropagator}),
it is straightforward to perform the necessary worldsheet integrals. 
In table 1  we give the prefactors, sums, and derivative 
structure that appear.  
\input{bosontable}

In the appendix we give some more details of this 
calculation. 
To order $8$ in derivatives we obtain a partition functional given by  
\beqn
Z\left( T(x) \right) &=& \exp\left[ - \tilde T(x) + \frac{\zeta(2)}{2}
{\alpha'}^2 \df_{\mu\nu} \tilde T(x) \df^{\mu\nu} \tilde T(x)
\right. \nonumber \\ && ~~~~
+ \zeta(3) {\alpha'}^3 
\df_{\mu\nu\gamma} \tilde T(x) \df^{\mu\nu\gamma} 
\tilde T(x) \nonumber \\ &&
~~~~+ \frac{1}{4} \left( \zeta^2(2) + 7 \zeta(4) \right) {\alpha'}^4 
\df_{\mu\nu\gamma\delta} \tilde T(x) \df^{\mu\nu\gamma\delta}
\tilde T(x) \nonumber \\ && 
~~~~- \frac{\zeta(3)}{3} {\alpha'}^3 \df_\mu^{~\nu} 
\tilde T(x) \df_\nu^{~\gamma} \tilde T(x) \df_\gamma^{~\mu} 
\tilde T(x)  \nonumber \\ &&
~~~~ - \frac{1}{2} \zeta^2(2) {\alpha'}^4 
\df_{\mu\nu\gamma\delta} \tilde 
T(x)
\df^{\mu\nu} \tilde T(x) \df^{\gamma\delta} \tilde T(x)
\nonumber \\
&& ~~~~
+ \left( \zeta^2(2) + \frac{1}{2} \zeta(4) \right) {\alpha'}^4
\df_{\mu\nu \gamma} \tilde T(x) \df^{\mu\nu}_{~~\delta} \tilde T(x)
\df^{\gamma \delta} \tilde T(x)
\nonumber \\
&& ~~~~ + \frac{1}{4} \zeta(4) {\alpha'}^4 \df_\mu^{~\nu}
\tilde T(x) \df_\nu^{~\gamma} \tilde T(x) \df_\gamma^{~\delta}
\tilde T(x) \df_\delta^{~\mu}
\tilde T(x)
\left. + \mathrm{O}(\df^{10}) \right].\nonumber \\
\label{eq:bosonicZ}
\eeqn
Here and throughout this paper whenever multiple derivatives act on the 
same field we abbreviate 
$\df_\mu \df_\nu \equiv \df_{\mu\nu}$.
Inspecting the expression (\ref{eq:bosonicZ}) we note that the 
terms 
$\frac{\zeta(n)}{n} 
(\df_{\mu}^{~\nu} \tilde T )^n$ that are familiar from the discussion of 
the 
quadratic 
tachyon profile \cite{Witten:1993cr,Kraus:2000nj,Laidlaw:2001jt} appear. 
In addition to this, the  
field redefinition (\ref{eq:nonlocalredef}) that is imposed by the 
world--sheet sigma model calculation gives a prescription for 
resolving some of the ambiguities in 
defining an off--shell tachyon action discussed in 
\cite{Fotopoulos:2003yt}.  
The substitution of $\tilde T$ 
for $T$ throughout reproduces that world--sheet cut--off dependent 
coefficient.
The Taylor expansion of (\ref{eq:nonlocalredef}) in powers of 
the cutoff--dependent coefficient exactly matches the first two terms of 
the 
renormalized field of \cite{Takayanagi:2000rz}.  
Equivalently, the partition functional  
(\ref{eq:bosonicZ}) can be expressed in terms of 
the original field $T(x)$ as
\beqn
Z\left( T(x) \right) &=& e^{-T} \left[ 1 + c \df_\mu^{~\mu} T - 
\frac{c^2}{2!} \df_{\mu\nu}^{~~\mu\nu} T + \frac{c^3}{3!} 
\df_{\mu\nu\alpha}^{~~~\mu\nu\alpha} T 
\right.
\nonumber \\
&& + \frac{c^2}{2!} \df_\mu^{~\mu} T \df_\nu^{~\nu} T
- \frac{c^3}{2!} \df_\mu^{~\mu} T \df_{\nu\alpha}^{~~\nu\alpha} T
+ \frac{c^3}{3!} \df_\mu^{~\mu} T \df_\nu^{~\nu} T \df_\alpha^{~\alpha} T
\nonumber \\
&& + \frac{\zeta(2)}{2}
{\alpha'}^2 \df_{\mu\nu}  T \df^{\mu\nu}  T
- \zeta(2) c {\alpha'}^2 \df_{\mu\nu\alpha}^{~~~\alpha} 
\df^{\mu\nu} T
\nonumber \\
&& \left. + \zeta(3) {\alpha'}^3
\df_{\mu\nu\gamma}  T \df^{\mu\nu\gamma}
 T
- \frac{\zeta(3)}{3} {\alpha'}^3 \df_\mu^{~\nu}
T \df_\nu^{~\gamma}  T \df_\gamma^{~\mu}
 T + \mathrm{O}(\df^8) \right].
\nonumber \\
\label{eq:bosonicZ2}
\eeqn
Using this expression, we could follow \cite{Tseytlin:2000mt} and fix the 
value of $c$ 
 to give 
a conventionally 
normalized 
kinetic term.  The expression 
in terms of $\tilde T$ (\ref{eq:bosonicZ}) does not 
contain any 
two--derivative terms, and appears more natural to us.  This situation is 
somewhat different than the 
superstring case, which is discussed in section \ref{sec:superfield}.

The term quadratic in the tachyon field (\ref{eq:grigt2}) can be 
Taylor expanded in $k_1 \cdot k_2$, giving
\beqn
\frac{\tilde T^2 }{2}  \left( 1 + \zeta(2) ( \alpha' k_1 \cdot k_2)^2 - 2 
\zeta(3) ( \alpha'  
k_1 
\cdot k_2)^3
+ \frac{19 \pi^2}{360} ( \alpha' k_1 \cdot k_2)^4 + O(k^{10}) \right), 
\eeqn
which exactly matches the terms quadratic in $\tilde T$ in the expansion 
of (\ref{eq:bosonicZ}).  In the case of the three point function, fixing 
two of the $k_{ij}$s in 
(\ref{eq:grigt3}) to be zero reduces it to (\ref{eq:grigt2}) times $\tilde 
T$, the terms in its expansion in a single $k_{ij}$ matches that of 
(\ref{eq:bosonicZ}).  Similarly, the expansion of (\ref{eq:grigt3}) to 
linear order in two of the $k_{ij}$ vanishes, while the term 
proportional to $k_{12} k_{13} k_{23}$ is $\frac{1}{3!}\psi''(1)= - 
\frac{1}{3} \zeta(3)$, in 
agreement with (\ref{eq:bosonicZ}).   The coefficient  proportional to 
$k_{12} 
k_{13} k_{23}^2$ is $\frac{3}{2} \psi'''(1)$ from expansion of 
(\ref{eq:grigt3}) which matches the $\frac{\pi^4}{30}$ from 
(\ref{eq:bosonicZ}).  
We emphasize that the analytic expressions obtained in 
(\ref{eq:grigt2}) and (\ref{eq:grigt3}) are valid for $k^2 > 0$, and that 
the perturbative 
expression is valid for $|k^2| \ll 1$.  
This fact that the derivative expansion is consistent with the results in 
two regimes, 
leads us to conjecture that both correctly capture the two--point and 
three--point 
dynamics of open string tachyons.  
We shall 
see that the situation for the two--point function 
is similar, but not exactly 
the same in the case of the superstring.

%% file: bosontable.tex
\begin{table}
\label{tbb} 
\begin{center}
\begin{tabular}{ c  c  c }
Prefactor & Derivative Form & Associated Sum \\
\hline $
-1 
$&$ 
T 
$&$
~
$\\$
\frac{1}{8} (2\alpha')^2
$&$
\df_{\mu\nu} T \df^{\mu\nu} T
$&$
\frac{1}{m^2}
$\\$
-\frac{1}{16} (2\alpha')^3
$&$
\df_{\mu\nu\alpha} T \df^{\mu\nu\alpha} T
$&$
\frac{1}{m n (m+n)}
$\\$
- \frac{1}{3 \cdot 4} (2\alpha')^3
$&$
\df_\mu^{~\nu} T \df_\nu^{~\alpha} T \df_\alpha^{~\mu} T
$&$
\frac{1}{m^3}
$\\$
\frac1{3! 16} (2\alpha')^4 
$&$
\df_{\mu\nu\alpha\beta} T \df^{\mu\nu\alpha\beta} T
$&$
\frac{4}{m n n' (m+n+n')} +  \frac{3 \delta_{m+m',n+n'}}{m m' n n'} 
$\\$
- \frac{1}{32} (2\alpha')^4
$&$
\df_{\mu\nu} T \df^{\mu}_{~\alpha \beta} T \df^{\nu \alpha \beta} T
$&$
\frac{2}{n^2 m (m+n)} + \frac{1}{m n (m+n)^2 }
$\\$
- \frac{1}{32} (2\alpha')^4
$&$
\df_{\mu\nu} T \df^{\mu\nu\alpha\beta} T \df_{\alpha \beta} T
$&$
\frac{1}{m^2 n^2}
$\\$
\frac{1}{32} (2\alpha')^4
$&$
\df_{\mu}^{~\nu} T \df_{\nu}^{~\alpha} T \df_\alpha^{~\beta} T
\df_\beta^{~\mu} T
$&$
\frac{1}{n^4}
$\\$
(-1)^k \frac{1}{k 2^{k-1}} (2\alpha')^k
$&$
\mathrm{Tr}(\df_\mu^{~\nu}T)^k
$&$
\frac{1}{m^k}
$
\end{tabular}
\end{center}
\caption{The terms which appear in the tachyon action for bosonic string 
theory around the constant 
tachyon, with their prefactors and the associated sums.  In all sums 
implicitly over the positive integers, and we use the abbreviation 
$\df_\mu \df_\nu \equiv \df_{\mu\nu}$ with generalizations to higher orders 
in derivatives.} \end{table}

%% file: superfield.tex
\section{Effective action for superstring tachyon}
\label{sec:superfield}

We use the intuition from the calculation in the bosonic sector to 
proceed
in the case of the superstring, and outline the calculation below.  Many
of the intermediate steps are detailed in the appendix.
We start with the conventions of \cite{Kutasov:2000aq}
for the action, using their superfield method to obtain the 
higher derivative terms for the tachyon partition function.  We define
superfields and a derivative by 
\beqn
\Gamma(\phi,\theta) &=& \eta(\phi)+\theta F(\theta),
\nonumber \\
{\bf X}^\mu(\phi,\theta)&=& X^\mu(\phi)+\theta\psi^\mu(\phi),
\nonumber \\
D&=& \partial_\theta+\theta\partial_\phi.
\eeqn
The boundary action coupled to a tachyon field is
\beqn
S_{boundary} =
\oint \cfrac{d \phi}{2\pi} d\theta [\Gamma D \Gamma + T({\bf X}) \Gamma].
\eeqn
The partition function is then given by 
\beqn
Z &=&\int[d{\bf X}][d\Gamma]e^{-S_{bulk}-S_{boundary}}
= \int[d{\bf X}][d\Gamma]e^{-S_0-S_1},
\label{eq:sustrZsec3}
\eeqn
with $S_{bulk}$ the standard NSR action in the bulk, and with the boundary 
superfield kinetic term included in $S_0$ giving
\beqn
S_0
&=& S_{bulk}+ \oint \cfrac{d\phi}{2\pi}d\theta \Gamma D \Gamma,
\nonumber \\
S_1 &=& \oint\cfrac{d\phi}{2\pi}T({\bf X})\Gamma.
\eeqn
In the superstring case, the tachyon action is equal to the partition 
function \cite{Kutasov:2000aq}.
As in the bosonic case we expand into a constant mode and fluctuations, 
with ${\bf X}=x+\tilde{\bf X}$, where $x$ denotes a constant mode
and $\tilde{X}$ encodes non-zero modes, and we do a Fourier transformation 
for the tachyon field
\beqn
T({\bf X})
&=& \int dk f(k) e^{i k \cdot {\bf X}}
=\int dk f(k) e^{i k \cdot x}
e^{i k \cdot \tilde{\bf X}},
\nonumber \\
S_1
&=& \int dk f(k) e^{i k \cdot x}
\oint\cfrac{d\phi}{2\pi}d\theta
e^{i k \cdot \tilde{\bf X}} \Gamma.
\eeqn
We can expand the partition function as
\beqn
Z = \int dx \int[d\tilde{\bf X}][d\Gamma]e^{-S_0-S_1}
&=&\int dx \int[d\tilde{\bf X}][d\Gamma]e^{-S_0}
\sum_{n=0}^{\infty}\cfrac{1}{n!}(-S_1)^n
\nonumber \\
&=& \int dx \sum_{n=0}^{\infty}\cfrac{1}{n!} \langle (-S_1)^n \rangle.
\eeqn
We note that since $S_1$ is proportional to $\Gamma$,
$\langle(-S_1)^n\rangle$ is non-vanishing only for even $n$, which allows 
us to write the partition function as
\beqn
Z = Z_0+Z_2+Z_4+Z_6+\ldots,
\eeqn
where the subscripts indicate the number of tachyon fields in each term. 
The $n$th order term is given by
\beqn
Z_n
&=& \cfrac{1}{n!} \int dx \langle (-S_1)^n \rangle
\nonumber \\
&=& \cfrac{1}{n!} \int dx \int \prod_{i=1}^n dk_i  f(k_i) 
e^{i (k_1 + \cdots k_n) \cdot x} \oint \prod_{i=1}^n \frac{d\phi_i}{2 \pi}
\nonumber \\
&&  \int d\theta_n \cdots d\theta_1
\langle e^{ik \cdot \tilde{\bf X}(1)} \cdots
e^{ik \cdot \tilde{\bf X}(n)}\rangle
\langle \Gamma(1) \cdots \Gamma(n) \rangle
\label{eq:superfieldZn}.
\eeqn
We have abbreviated the $\phi_i,~\theta_i$ dependence by making their 
subscript $i$ the argument.  The superfield propagators are
\beqn
\langle\tilde{\bf X}(\phi_1,\theta_1)\tilde{\bf X}(\phi_2,\theta_2)\rangle
&=& \langle\tilde{X}(\phi_1)\tilde{X}(\phi_2)\rangle
-\theta_1\theta_2\langle\psi(\phi_1)\psi(\phi_2)\rangle,
\nonumber \\
\langle\Gamma(\phi_1,\theta_1)\Gamma(\phi_2,\theta_2)\rangle
&=& \langle \eta(\phi_1)\eta(\phi_2)\rangle+\theta_1\theta_2
\langle F(\phi_1)F(\phi_2)\rangle,
\eeqn
with boundary-to-boundary  propagators for the component fields given by 
\beqn
\langle\tilde{X}(\phi_1)\tilde{X}(\phi_2)\rangle
&=& \alpha' \sum_{{\scriptstyle m\in \mathbb{Z}} \atop{\scriptstyle 
m\not=0}}
\cfrac{1}{|m|}e^{im(\phi_1-\phi_2)},
\nonumber \\
\langle\psi(\phi_1)\psi(\phi_2)\rangle
&=& \alpha' i\sum_{r\in \mathbb{Z}+\frac{1}{2}}
\cfrac{r}{|r|}e^{ir(\phi_1-\phi_2)},
\nonumber \\
\langle F(\phi_1)F(\phi_2)\rangle
&=&
\cfrac{1}{2}\sum_{m\in \mathbb{Z}}e^{im(\phi_1-\phi_2)},
\nonumber \\
\langle\eta(\phi_1)\eta(\phi_2)\rangle
&=&
\cfrac{1}{2}i \sum_{r \in \mathbb{Z}+\frac{1}{2}}
\cfrac{1}{r}e^{ir(\phi_1-\phi_2)}.
\eeqn
As in the bosonic case, the propagator at zero separation for the 
field $\tilde X$ is divergent.   The propagator of $F(\phi)$ is a 
$\delta$-function
$\pi \delta(\phi_1-\phi_2)$ for periodic functions and the $\eta$ 
propagator is a step function.
We can evaluate 
(\ref{eq:superfieldZn}) and obtain the $n$--point function 
\beqn
\langle
e^{ik_1\cdot\tilde{\bf X}(1)} \ldots e^{ik_n\cdot\tilde{\bf X}(n)}
\rangle
&=& e^{-({k_1}^2+\ldots+{k_n}^2) c}
\prod_{i<j=1}^n e^{- k_i \cdot k_j
\langle \tilde{\bf X}(i)\tilde{\bf X}(j)\rangle} \label{eq:sustrnptx}\\
\langle \Gamma(1) \ldots \Gamma(n) \rangle &=& \sum_{\cal{P}} 
(-1)^{\cal{P}}  \langle \Gamma({\cal{P}}(1)) 
\Gamma({\cal{P}}(2)) \rangle \nonumber \\&& ~~~~~~~~~~~\ldots \langle 
\Gamma({\cal{P}}(n-1)) 
\Gamma({\cal{P}}(n) ) \rangle 
\label{eq:sustrmptgamma}
\eeqn
The sum in (\ref{eq:sustrmptgamma}) is over the $(n-1)!!$ pairwise 
distinct 
permutations of the indices $1 \ldots n$.
The $c$ in (\ref{eq:sustrnptx}) is the same as that in 
(\ref{eq:nonlocalredef}), and 
those terms will be used rescale $T$ to $\tilde T$.  Substituting 
(\ref{eq:sustrnptx}) and (\ref{eq:sustrmptgamma}) into 
(\ref{eq:superfieldZn}) and integrating over the Grassman coordinates is  
equivalent to using the derived
expression 
for the boundary interaction that was 
obtained by
taking (\ref{eq:sustrZsec3}) and integrating out the boundary superfield.
The action given by 
\cite{Kutasov:2000aq} 
is
\beqn
S &=& \frac{1}{ 2 \alpha'} \int \frac{d^2 z}{2\pi} \left( \df X^\mu \ddf
X_\mu + \psi^\mu \ddf \psi_\mu + \tilde \psi^\mu \df \tilde \psi_\mu
\right)
 + \frac{1}{4} \oint \frac{d\phi}{2\pi} T^2\left(X(\phi)\right)
\nonumber \\
&&
+ \frac{1}{4} \oint \frac{d\phi d\phi'}{4 \pi}
\epsilon(\phi-\phi')
 \Big(\psi^\mu(\phi) \df_\mu T(X(\phi)) \Big)
\Big(\psi^\mu(\phi') \df_\mu T(X(\phi')) \Big),
\nonumber \\
\label{eq:sustraction}
\eeqn
where the function $\epsilon(x)$ is the sign function,
so that $\epsilon(x) =1$ for $x>0$ and $\epsilon(x)=-1$ for $x<0$.  As we 
consider the 
NS-sector, this relates  $\psi$ and $\tilde \psi$  on 
the 
boundary.
This interactions features both a non--local fermion interaction and a 
contact term
proportional to $T^2$.

There are two well known exact expressions for the superstring tachyon for   
tachyons with a single momentum mode like 
those
discussed in the context of the bosonic string.  The first is the expansion 
around the constant tachyon, which gives rise to a potential 
$e^{-\frac{\tilde T^2}{4} }$ \cite{Kutasov:2000aq}, and the second is the 
expansion 
around the rolling tachyon with $k^2=\frac{1}{2}$ which has a
potential of $\frac{1}{1+ \frac{1}{2} \tilde T^2 }$ \cite{Larsen:2002wc}.  
In fact, 
for any momentum, the term bilinear in the tachyon couplings can be written 
as
\beqn
\langle \frac{1}{2!} \oint T(\phi_1) T(\phi_2) \rangle
&=&
\frac{1}{2}\int dk_1dk_2 f(k_1)f(k_2)e^{i(k_1+k_2)\cdot x}
e^{-\frac{1}{2}(k_1^2+k_2^2)\delta}
\nonumber \\
&&
\times
\oint \cfrac{d\phi_1}{2\pi}\cfrac{d\phi_2}{2\pi}d\theta_2 d\theta_1
\langle \Gamma(1)\Gamma(2) \rangle
e^{-k_1\cdot k_2\langle \tilde{\bf X}(1)\tilde{\bf X}(2) \rangle }.
\eeqn
The momentum dependence is encoded in the integral 
\beqn
\oint \cfrac{d\phi_1}{2\pi}\cfrac{d\phi_2}{2\pi}d\theta_2d\theta_1
\Big(
\langle \eta(1)\eta(2)\rangle
+\theta_1\theta_2\langle F(1)F(2)\rangle
\Big)
\nonumber \\
 \times e^{- \alpha' k_1\cdot k_2 \left( \langle 
\tilde{X}(1)\tilde{X}(2)\rangle
-\theta_1 \theta_2 \langle \psi(1)\psi(2) \rangle \right)}
&=& 
\nonumber \\ 
\oint \frac{d \phi_1}{2 \pi} \frac{d\phi_2}{2\pi} \bigg(
\pi \delta(\phi_1-\phi_2) + 2 \alpha' k_1\cdot k_2 \frac{  \epsilon(
\phi_1-\phi_2)
}{ 2 \sin \frac{ \phi_1-\phi_2}{2} } \bigg)
&
 \times& \left| 2 \sin \frac{ \phi_1-\phi_2}{2} \right|^{2\alpha'
k_1\cdot
k_2}.
\label{eq:sustrpropagator}
\eeqn
For $k_1\cdot k_2 >0$, the contact term in (\ref{eq:sustrpropagator}) 
vanishes, and we are left with 
\beqn
\langle \frac{1}{2!} \oint T(\phi_1) T(\phi_2) \rangle
&=&
\frac{\tilde T^2}{2!} \pi \frac{ \Gamma(1+2 \alpha' k_1\cdot k_2) }{ 
\Gamma^2(\frac{1}{2} + \alpha' k_1\cdot k_2) }
\eeqn
while for $k_1\cdot k_2 <0$ both integrals diverge and must be defined by 
regularization, as we discuss below. 

In analogy with the calculation outlined for the bosonic string we 
can calculate an expansion in powers of momentum around the 
background of a constant tachyon.  The expansion parameter is the 
momentum, which is presumed to be small, so we  
 expand the exponential in $k$, which will 
be Fourier transformed into powers of derivatives as before.  In table 
2 we list the sums and derivative structures that appear 
in the Taylor expansion of (\ref{eq:sustrZsec3}) about $k=0$.

%% file: superstring.tex

From the explicit formulae for $Z_n$, we can extract an expression for the 
partition functional.  
After some manipulations,  the 
expression for the partition functional of 
$\tilde T$ is given as
\beqn
Z\left(\tilde T(x) \right) &=& \exp \left( -\frac{1}{4} {\tilde T}^2
+ \alpha' \ln 2 \df^\mu \tilde T \df_\mu \tilde T
+ {\alpha'}^2 \left( \frac{1}{2} \zeta(2) -2 (\ln 2)^2 \right) 
\df^{\mu\nu} \tilde T \df_{\mu\nu} \tilde T
\right. \nonumber \\ &&
+ {\alpha'}^3 \left( \frac{1}{2} \zeta(3) -4 \zeta(2) \ln 2 +
 \frac{8}{3} (\ln 2)^3 \right) \df^{\mu\nu\alpha}
\tilde T \df_{\mu\nu \alpha} \tilde T
\nonumber \\ &&
+ {\alpha'}^2 \frac{1}{2^5} \zeta(2) \df^{\mu\nu} {\tilde 
T}^2
\df_{\mu\nu} {\tilde T}^2 
- {\alpha'}^2 \frac{3}{8} \zeta(2) \df^\mu \tilde T \df_\mu \tilde T
\df^\nu \tilde T \df_\nu \tilde T 
\nonumber \\ &&
+ {\alpha'}^3 \frac{1}{2^4} \zeta(3) \df^{\mu\nu\alpha}
{\tilde T}^2
 \df_{\mu\nu \alpha} {\tilde T}^2 - {\alpha'}^3 \frac{1}{2^4} \zeta(2) 
\ln 2 
\df^\mu \tilde T \df_{\mu\nu \alpha} \tilde T \df^{\nu\alpha} {\tilde
T}^2 \nonumber \\ && 
+ {\alpha'}^3 \frac{1}{8} \left( 7 \zeta(3) - 4 \ln 2
\right) \df^{\mu\nu} \tilde
T \df^\alpha_{~\mu} \tilde T \df_{\nu\alpha} {\tilde T}^2 
\nonumber \\ &&
+ {\alpha'}^3 \frac{7}{4} \zeta(3) \df^\mu \tilde T \df_{\mu\nu} \tilde T
\df^{\nu\alpha} \tilde T \df_\alpha \tilde T - {\alpha'}^3 \frac{1}{2^5 3}
\zeta(3) \df_{\mu\nu} {\tilde T}^2 \df^{\nu\alpha} {\tilde T}^2
\df_\alpha^{~\mu} {\tilde T}^2 
\nonumber \\ && 
+ {\alpha'}^3 \frac{1}{4}
\left( -8 \zeta(2) \ln 2 + 14 \zeta(3) \right) \df^{\mu\nu} \tilde T
\df_{\mu\nu} \tilde T \df^\alpha \tilde T \df_\alpha \tilde T 
\nonumber \\ &&
\left. + {\alpha'}^3 \frac{7}{2^3 3} \zeta(3) \df^\mu \tilde T \df_\mu
\tilde T \df^\nu \tilde T \df_\nu \tilde T \df^\alpha \tilde T \df_\alpha
\tilde T + \mathrm{O} (\df^8) \right). 
\label{eq:sustrZ} 
\eeqn 
The coefficients that appear in (\ref{eq:sustrZ}) have been regularized as 
in the  appendix.
Since they diverge individually, it is possible to obtain any finite 
value for the difference between the sums that come from the worldsheet
bosons and fermions.
The choice of  
$\zeta$--function regularization, which is equivalent 
to a choice of field redefinition, fixes these coefficients.  Unlike 
(\ref{eq:bosonicZ}) there is term that is
quadratic in both derivatives and the field $\tilde T$, a standard kinetic 
term, however unusual normalization is an artifact of this 
$\zeta$--function regularization procedure. 
  As in the bosonic case it is 
possible to extend this analysis to higher numbers of derivatives.
\input{sustrtable}

We discussed earlier the expression exact in $k_1 \cdot k_2>0$ for the two 
point function.  For $k^2 \ll 1$ the relevant expansion can be expressed 
as a difference of divergent sums.  
The higher derivative terms 
that are quadratic in $\tilde
T$ can be derived or read off from table 2  and the 
`kinetic' term 
with 
order $\df^{2n}$ derivatives is given by the regularization of 
\beqn
-\frac{1}{4} \frac{1}{n!} \df^{\mu_1 \ldots \mu_n} \tilde T
\df_{\mu_1 \ldots \mu_n} \tilde T \sum
\left[ \frac{1}{m_1 \ldots m_n} - \frac{n}{2^n} \frac{1}{m_1 \ldots 
m_{n-1} } \frac{1}{r \pm m_1 \ldots \pm m_{n-1} } \right].
\nonumber \\
\eeqn
In this, the sum is taken over the positive integers $m$ and 
half--integers $r$, and there is an implicit sum over all possible signs 
in the second denominator following the conventions of the appendix.
The first sum can be seen as a manifestation of the divergent 
term in (\ref{eq:sustrpropagator}), while the second comes from the 
world--sheet fermions.
Naively, both sets are divergent proportional to $\ln \Lambda$, where 
$\Lambda$ is the cut--off of the world--sheet theory.   
Using our conventions, 
the first two terms in the derivative expansion that are quadratic 
in
$\tilde 
T$ are
\beqn
-\frac{1}{4} {\tilde T}^2 + \alpha' \ln2 \df^\mu \tilde T \df_\mu \tilde 
T
\eeqn
in agreement with \cite{Tseytlin:2000mt}.

%% file: sustrtable.tex
\begin{table}[ht]
\begin{center}
\begin{tabular}{ r  c  c }
Prefactor & Derivative Form & Associated Sum \\
\hline $
-\frac14 
$&$ 
T^2 
$&$
~
$\\$
- \frac14 2\alpha' 
$&$ 
\df^\mu T \df_\mu T 
$&$ 
\left(\frac1n - \frac1r \right) 
$\\$
-\frac18 (2\alpha')^2
$&$
\df_{\mu\nu} T \df^{\mu\nu} T
$&$
\left( \frac1n
\frac1m - \frac1n \frac1{r\pm n}  \right)
$\\$
\frac{1}{4^2 2^3} (2 \alpha')^2
$&$
\df_{\mu\nu} T^2 \df^{\mu\nu} T^2
$&$
\frac{1}{n^2}
$\\$
-\frac{1}{4^2 2} (2\alpha')^2
$&$
\df_\mu T \df^\mu T \df_\nu T \df^\nu T
$&$
\frac{1}{r^2}
$\\$
-\frac{1}{4 \cdot 3!} (2\alpha')^3
$&$
\df_{\mu\nu \alpha} T \df^{\mu\nu \alpha} T
$&$
\left( \frac{1}{n n' m} - \frac{3}{2^2}\frac{1}{ n n'}  \frac1{r \pm n \pm 
n'}
\right)
$\\$
\frac{1}{4^2 2!^2 4} (2\alpha')^3
$&$
\df_{\mu\nu \alpha} T^2
\df^{\mu\nu \alpha} T^2
$&$
\frac{1}{n n' (n+n')}
$\\$
\frac{1}{4^2 2!^2} (2\alpha')^3
$&$
\df_{\mu\nu} \left( \df_\alpha T
\df^\alpha T \right) \df^{\mu\nu} T^2
$&$
\frac{1}{n^2 n'}
$\\$
- \frac{1}{4^2 4} (2 \alpha')^3
$&$
\df_{\mu\nu} T \df^\mu_{~\alpha} T
\df^{\nu\alpha} T^2
$&$
\frac{1}{n^2}  \frac{1}{r \pm n}
$\\$
- \frac{1}{4^2 2! } (2 \alpha')^3
$&$
\df_\mu T \df^{\mu\nu\alpha} T \df_{\nu
\alpha} T^2
$&$
\frac{1}{r n^2}
$\\$
- \frac{1}{4^2 2!} (2 \alpha')^3
$&$
\df_\mu T \df^{\mu\nu} T \df_{\nu\alpha}
T
\df^\alpha T
$&$
\left( \frac{1}{r n} ( \frac{1}{r-n} + \frac{3}{r+n} )
+ \frac{1}{r r' (r+r')} \right)
$\\$
- \frac{1}{4^2 2! } (2 \alpha')^3
$&$
\df_{\mu\nu} T \df^{\mu\nu} T \df_\alpha T \df^\alpha T
$&$
\left( \frac{1 }{n r^2} + \frac{1}{n (n+r)^2} - \frac{1}{ r^2 (r+r')}
\right)
$\\$
-\frac{1}{4^3 3! 2} (2 \alpha')^3
$&$
\df_{\mu\nu} T^2 \df^{\nu \alpha} T^2
\df_{\alpha}^{~\mu}T^2
$&$
\frac{1}{n^3}
$\\$
\frac{1}{4^3 3} (2 \alpha')^3
$&$
\df^\mu T \df_\mu T \df^\nu T \df_\nu T
\df^\alpha T \df_\alpha T
$&$
\frac{1}{r^3}
$
\label{table:sustr}
\end{tabular}
\end{center}
\caption{The terms that appear in the tachyon action for 
superstring theory in the expansion around the constant tachyon, 
with prefactors and sums.  The sums are over positive integers for 
$n,m\ldots$ and positive half integers for $r,s \ldots$.  The terms 
$\frac{1}{r\pm n \pm m}$ implicitly sum over all possible combinations of 
signs.} \end{table}

%% file: comparisons.tex
Having calculated an action in the derivative expansion for the 
superstring tachyon out to $6$ orders in 
derivatives, we  check for consistency with
other expressions for tachyon actions that have been derived previously.
We start with the observation that in the exactly integrable cases 
discussed \cite{Witten:1993cr,Craps:2001jp,Kutasov:2000aq}, namely the 
quadratic and linear tachyon profiles  for the bosonic and superstring 
cases respectively. We 
reproduce the first terms of the exact action, as expected since 
we   reproduce the techniques that were used to derive these initially.

Similarly, we can compare with the calculation \cite{Takayanagi:2000rz} 
which  expands around the constant 
tachyon 
background.  Taking their expression for the tachyon effective action to 
order ${\alpha'}^2$ in the absence of gauge fields and for a real tachyon, 
we have
\beqn
Z
&=&
T_9 e^{-T^2} \Big[ 
2+8\alpha'\ln 2 \left(\partial_\mu T\right)^2
+4\alpha'^2 \gamma_0 \left(\partial_\mu \partial_\nu T\right)^2
\nonumber \\ && +4\alpha'^2\left(
4\left(\ln 2\right)^2-\zeta(2) 
\right)
\left(\partial_\mu T\partial^\mu T\right)^2
-\alpha'^2\cfrac{2\pi^2}{3}\left(\partial_\mu T\partial^\mu T\right)^2
\nonumber \\ && 
 +2 \zeta(2) \alpha'^2
T \partial^\mu \partial^\nu T 
\left(
\partial_\mu \partial_\nu T^2
+2 \partial_\mu \partial_\nu T
\right)
\Big].
\label{eq:TTUz}
\eeqn
Here $T_9$ is the D9 brane tension, $T$ is the renormalized tachyon 
field of \cite{Takayanagi:2000rz}, and the parameter 
$\gamma_0$ is defined in the small $\epsilon$ limit as the regularized sum
\beqn
\gamma_0
 &=&\sum_{m=1}^\infty \sum_{r=\frac{1}{2}}^\infty\cfrac{1}{m}
\left(
\cfrac{1}{r+m}+\cfrac{1}{r-m}
\right)
e^{-(r+m)\epsilon} - (\ln \epsilon)^2.
\label{eq:gamma}
\eeqn
To show the correspondence with our results (\ref{eq:sustrZ}), we  
first rewrite
\beqn
(\ln \epsilon)^2 = \sum_{m,n=1}^\infty \frac{1}{m n} e^{-\epsilon (m + n)}
= \sum_{m,n=1}^\infty \frac{2}{m (m+n)} e^{-\epsilon (m + n)}
\eeqn
and then see that in the $\epsilon 
\rightarrow 0$ limit (\ref{eq:gamma}) is our sum (\ref{eq:4ktermapp}) 
and so 
\beqn
\gamma_0
&=& - 4 (\ln 2)^2 + 2 \zeta(2)
\eeqn
as  
detailed in the appendix.  Using this regularization, the terms from 
(\ref{eq:TTUz}) match those that we obtain.  In addition, the off--shell 
field 
redefinition (\ref{eq:nonlocalredef}) is seen in the renormalized field 
$T_R$ in \cite{Takayanagi:2000rz}.
The
field redefinition is ubiquitous, and so appears in any kind of string 
effective action, not only for tachyons, but also for massless and massive 
modes.  This is a rescaling of the boundary fields on--shell. 
Whether the rescaling enhances or suppresses the field is determined by 
the field's worldsheet dimension (relevance, irrelevance or marginality).

%% file: concl.tex
\section{Discussion and Conclusions}
\label{sec:concl}

In this note, we have examined some of the issues associated with the 
derivative expansion of the (super)string tachyon action.  In particular, 
we 
wished to focus on the compatibility of the derivative expansion in 
different 
momentum 
regimes.  To this end, we have given an expression for the partition 
function of the bosonic string coupled to a single tachyon momentum mode 
(\ref{eq:bosonplanewave}).  This interpolates between the well--known 
results for the 
constant and rolling tachyons \cite{Witten:1993cr,Larsen:2002wc}.  This 
also reproduces the spatial
derivative expansion around the rolling tachyon derived by
\cite{Fotopoulos:2003yt,Kutasov:2003er}.  
Our expansion 
around the 
constant tachyon (\ref{eq:bosonicZ}) reproduces the first terms of 
an exact result for the three tachyon coupling calculated by 
\cite{Coletti:2004ri}.  
We have identified a field redefinition (\ref{eq:nonlocalredef}) which 
reproduces that of \cite{Fotopoulos:2003yt, Takayanagi:2000rz} for the 
world--sheet cut--off 
dependent coefficients in the tachyon action.  This has been noticed by 
other authors \cite{Coletti:2004ri,Fujita:2004ha}, however, we believe it 
worth remarking on 
because of the prevalence in the literature of this kind of term.  For the 
case of the superstring tachyon, we have extended by two orders in 
derivatives the calculations 
of \cite{Takayanagi:2000rz}.  Our calculation includes a hint 
about the 
purpose of the contact terms in the worldsheet theory.  These contact terms 
are necessary for unbroken world--sheet supersymmetry and contribute to 
some of the originally calculated amplitudes \cite{Kraus:2000nj}, but 
are absent in the context of calculations around the rolling tachyon.  The 
reason for this seems to be that the tachyon correlation functions are 
similar to  
$\beta$ functions, ratios of $\Gamma$ functions.  These have a particular 
domain of convergence for 
the integrals involved, and can be analytically continued to other regions.  
The contact terms seem to act to regularize the integrals that arise, and 
here we have presented the circumstantial evidence that the terms in the 
derivative expansion of the tachyon effective action, while some are 
naively divergent, admit a $\zeta$--function  regularization 
which matches with exact results.

\subsection*{Acknowledgements}

The authors would like to thank R. Myers and N.
Suryanarayana for discussions on this project, and A. Tseytlin for a 
clarifying discussion on \cite{Fotopoulos:2003yt}.  
MN would like to thank 
K. Hashimoto, K. Hosomichi, S. Terashima for 
useful discussions, and Y. Schroeder for advice on calculational 
techniques.  ML's research is supported by 
NSERC 
Canada.

%% file: appSums.tex
\section{Sums and Regularization Methods}
\label{sec:sumsappendix}

Here we tabulate some useful sums that appear in the world--sheet approach 
to this problem.  Throughout this we will use the convention that $m, n 
\ldots$ 
are positive integers and $r,s \ldots$ are positive half--integers.
We use a $\zeta$-function regularization, with
 the definition of Riemann's $\zeta$-function 
\beqn
\zeta(z) = \sum_{n=1}^\infty \frac{1}{n^z}.
\eeqn
This has a pole at $z=1$, and an analytic continuation throughout the 
complex $z$ plane.   
This can also be related to a sum over the positive half-integers, as
\beqn
\sum_{r=1/2}^\infty \frac{1}{r^z} &=& \sum_{m=1}^\infty 
\frac{2^z}{(2m-1)^z} + \sum_{m=1}^\infty \frac{2^z}{(2m)^z} - \zeta(z)
\nonumber \\
&=& \left( 2^z-1 \right) \zeta(z).
\eeqn
A related function which will be useful for our 
purposes is the $\psi$ function, defined as
\beqn
\psi(x) = \frac{d}{dx} \ln \Gamma(x),
\eeqn
with the series representation
\beqn
\psi(x) = -\gamma -\frac{1}{x} + x \sum_{k=1}^\infty \frac{1}{k (x+k)},
\eeqn
where $\gamma$ is Euler's constant, $\gamma \approx 0.5772$.  The 
derivatives of $\psi(x)$ satisfy
\beqn
x \psi^{(n)}(x) = (-1)^{n+1} n! \sum_{k=1}^\infty \frac{1}{(x+k)^{n+1} }.
\eeqn

We also note that with the help of the integral $\int_0^1 x^m (\ln x)^n dx 
= 
\frac{n!}{(m+1)^{n+1}}$
the  
sums of the form 
\beqn
\sum_{m_1 \ldots m_n =1}^\infty \frac{1}{m_1 \ldots m_n (m_1 + \ldots m_n) 
}
&=& n! \zeta(n+1)
\eeqn
can be evaluated.

In table 1 we list two sums that are not covered by the 
above considerations
\beqn
 \sum \frac{4}{ m n n' (m+n +n')} + 
\frac{3}{m m' n n'} \delta_{m+m',n+n'}
&=&  8 \oint \frac{d\phi}{2\pi} |\ln( 2 \sin\frac{\phi}{2} )|^4
\nonumber\\
&=& 6 \left( \zeta(2)^2 + 7 \zeta(4) \right),
\\
\sum \frac{2}{n^2 m (m+n) } + \frac{1}{m n (m+n)^2} &=& \zeta(2)^2 + 2 \sum 
\frac{1}{m (m+n)^3} 
\nonumber \\
&=& \zeta(2)^2 + \frac{1}{2} \zeta(4).
\eeqn

For the superstring, we give explicitly our regularizations of the 
sums that appear in the expansion of the tachyon bilinear term
in powers of momentum.
\beqn
\sum \frac{1}{n} - \sum \frac{1}{r} &=& \sum \left( \frac{1}{m} - 
\frac{1}{m-\frac12} \right)
\nonumber \\
&=& \lim_{x \rightarrow 1} 2 \sum_{n=1}^\infty \frac{(-x)^n}{n} 
\nonumber \\
&=& -2 \ln 2
\eeqn
This can alternatively be obtained using representations of the infinite 
sums as logarithmic derivatives of the $\Gamma$--function.  In the case of 
$4$ derivatives we find 
\beqn
\sum \left(\frac{1}{m n} - \frac{1}{m(r\pm m)} \right) 
\nonumber &=& 
\sum \left( \frac{2}{m (n+m) } -\frac{1}{m(r+m)} - \frac{1}{m(r-m)} 
\right)
\nonumber \\
&=& - \sum \frac{2}{m} \left( \psi(m+1) - \psi( m+ \frac12) \right)
\nonumber \\
&=& - 4 \lim_{x \rightarrow 1} \sum \frac{x^{2m}}{m} \left( \ln 2 
+ \sum_{k=1}^{m}
\frac{1}{2k} - \frac{1}{2k-1}  \right)
\nonumber \\
&=& 
- 4 \lim_{x \rightarrow 1} \sum \frac{x^{2m}}{m} \left( \ln 2 + 
\frac{1}{2m} + \sum_{k=1}^{2m-1} \frac{(-1)^k}{k} \right)
\nonumber \\
&=& -2 \zeta(2) + 4 (\ln 2)^2
\label{eq:4ktermapp}
\eeqn
We have abbreviated $\frac{1}{m (r\pm m)} \equiv \frac{1}{m (r+m)} + 
\frac{1}{m (r-m)}$ as mentioned in table 2.
In (\ref{eq:4ktermapp}) we have used the symmetry of the $\psi$, namely, 
$\psi( n + 
\frac{1}{2}) = \psi(\frac{1}{2} - n)$ for $n \in \mathbb Z$, and the 
identity $\ln (1-x) \ln(1+x) = \sum_k \frac{x^{2k}}{k} \sum_{n=1}^{2k-1} 
\frac{(-1)^n}{n} $.
In the 6 derivative case we find
\beqn
\sum \frac{1}{n n' m} - \frac{3}{2^2} \frac{1}{n m} \frac{1}{r \pm n \pm 
m} &=& 
3 \sum \frac{1}{n m (n' + m + n)} - \frac{1}{4} \frac{1}{n m (r \pm n \pm 
m)}
\nonumber \\
&=& 3 \sum \frac{1}{m n} \psi(1+m+n) - \frac{1}{4} \psi(\frac{1}{2} \pm n 
\pm m)
\nonumber \\
&=& 3 \sum_{m,n=1}^\infty \frac{1}{m n} \left( 2 \ln 2 + \sum_{k=1}^{m+n} 
\frac{1}{k} \right. \nonumber \\ && ~~~~\left. 
- \sum_{k=1}^{m+n} \frac{1}{2k-1} - \sum_{k=1}^{|m-n|} \frac{1}{2k-1}
\right).   
\eeqn
In this sum, the third and fourth term can be related by the observations 
that
\beqn
\sum_{n,p=1}^\infty\cfrac{1}{np}\sum_{m=1}^{n+p}\cfrac{1}{2m-1}
&=& \sum_{n=1}^\infty\sum_{l=n+1}^\infty
\cfrac{1}{n(l-n)}\sum_{m=1}^l\cfrac{1}{2m-1}
\nonumber \\
&=& 2\sum_{l=1}^\infty
\left( \sum_{n=1}^l \cfrac{1}{ln}-\cfrac{1}{l^2} \right)
\sum_{m=1}^l\cfrac{1}{2m-1},
\eeqn
and 
\beqn
\sum_{n,p=1}^\infty \frac{1}{np} \sum_{m=1}^{|n-p|}\cfrac{1}{2m-1}
&=& 2\sum_{{\scriptstyle n,p=1} \atop{\scriptstyle n>p}}^\infty
\cfrac{1}{np}\sum_{m=1}^{n-p}\cfrac{1}{2m-1}
\nonumber \\
&=& 2\sum_{n=2}^\infty\sum_{l=1}^{n-1}\cfrac{1}{n(n-l)}
\sum_{m=1}^{l}\cfrac{1}{2m-1}
\nonumber \\
&=& 2\sum_{l=1}^\infty\cfrac{1}{l}\sum_{n=1}^l \cfrac{1}{n}
\sum_{m=1}^l \cfrac{1}{2m-1},
\eeqn
from which we see that the two sums are related as
\begin{equation}
\sum_{n,p=1}^\infty\cfrac{1}{np}
\sum_{m=1}^{|n-p|}\cfrac{1}{2m-1}
=\sum_{n,p=1}^\infty\cfrac{1}{np}
\sum_{m=1}^{n+p}\cfrac{1}{2m-1}
+2\sum_{l=1}^\infty\cfrac{1}{l^2}\sum_{m=1}^l\cfrac{1}{2m-1}.
\end{equation}
This allows us to regularize
\beqn
\sum \frac{1}{n n' m} - \frac{3}{2^2} \frac{1}{n m} \frac{1}{r \pm n \pm
m} &=& 3 \sum \frac{2}{m n} \left( \sum_{k=1}^{2(n+m)} \frac{(-1)^k}{k} + 
2\ln 2 \right) 
\nonumber \\ && ~~~~~~~~
-6 \sum \frac{1}{m^2} \sum_{k=1}^{m} \frac{1}{2k 
-1}
\nonumber \\ 
&=& 6 \int_0^1 dx \left( \frac{\ln(1-x^2)^2}{1+x} 
- \frac{\ln(1-x^2)^2}{2x} \right.
\nonumber \\
&&~~~~~~\left. 
+ 2 \frac{\ln(1+x) 
\ln(1-x)}{x}  \right)
\nonumber \\
&=& 8 (\ln 2)^3 + \frac32 \zeta(3) - 12 \zeta(2) \ln 2.   
\eeqn

Using similar manipulations we can obtain:
\beqn
\sum \frac{1}{n n' (n+n')} &=& 2 \zeta(3)
\\
\sum \frac{1}{n (n+n')^2} &=& \zeta(3)
\\
\sum \frac{1}{r r' (r+r')} &=& 7 \zeta(3)
\\
\sum \frac{1}{n (n+r)^2} &=& -6 \zeta(2) \ln 2 + 7 \zeta(3)
\\
\sum \frac{1}{r (n+r)^2} &=& 6 \zeta(2) \ln 2 - \frac{7}{2} \zeta(3)
\\
\sum \frac{1}{m^2} \left( \frac{1}{n}  - \frac{1}{2} \frac{1}{r \pm m}  
\right) &=& \frac{7}{2} \zeta(3) - 2 \zeta(2) \ln 2
\\
\sum \frac{1}{r n (r+n)} &=& \frac{7}{2} \zeta(3)
\\
\sum \frac{1}{r n} \left( \frac{1}{r+n} - \frac{1}{r-n} \right) &=& 7 
\zeta(3)
\\
\sum \frac{1}{r^2} \left( \frac{1}{n} - \frac{1}{r+r'} \right) &=& 
-6 \zeta(2) \ln 2 + 7 \zeta(3).
\eeqn